\documentclass[twocolumn,showpacs,preprintnumbers,amsmath,amssymb,superscriptaddress]{revtex4}

\usepackage{graphicx}
\begin{document}
\title{Hole-doping-induced changes in the electronic structure 
of La$_{1-x}$Sr$_x$FeO$_3$ : soft x-ray photoemission and
absorption study of epitaxial thin films}

\author{H. Wadati}
\affiliation{Department of Physics and Department of Complexity 
Science and Engineering, University of Tokyo, 
Bunkyo-ku, Tokyo 113-0033, Japan}
\author{D. Kobayashi}
\affiliation{Department of Applied Chemistry, University of Tokyo, 
Bunkyo-ku, Tokyo 113-8656, Japan}
\author{H. Kumigashira}
\affiliation{Department of Applied Chemistry, University of Tokyo, 
Bunkyo-ku, Tokyo 113-8656, Japan}
\author{K. Okazaki}
\affiliation{Department of Physics and Department of Complexity 
Science and Engineering, University of Tokyo, 
Bunkyo-ku, Tokyo 113-0033, Japan}
\author{T. Mizokawa}
\affiliation{Department of Physics and Department of Complexity 
Science and Engineering, University of Tokyo, 
Bunkyo-ku, Tokyo 113-0033, Japan}
\author{A. Fujimori}
\affiliation{Department of Physics and Department of Complexity 
Science and Engineering, University of Tokyo, 
Bunkyo-ku, Tokyo 113-0033, Japan}
\author{K. Horiba}
\affiliation{Department of Applied Chemistry, University of Tokyo, 
Bunkyo-ku, Tokyo 113-8656, Japan}
\author{M. Oshima}
\affiliation{Department of Applied Chemistry, University of Tokyo, 
Bunkyo-ku, Tokyo 113-8656, Japan}
\author{N. Hamada}
\affiliation{Department of Physics, Tokyo University of Science, 
2641 Yamazaki, Noda, Chiba 278-8510, Japan}
\author{M. Lippmaa}
\affiliation{Institute for Solid State Physics, University of Tokyo, 
Kashiwanoha 5-1-5, Kashiwa, Chiba 277-8581, Japan}
\author{M. Kawasaki}
\affiliation{Institute for Materials Research, Tohoku University, 
2-1-1 Katahira, Aoba, Sendai 980-8577, Japan}
\author{H. Koinuma}
\affiliation{Materials and Structures Laboratory, Tokyo Institute of Technology
4259 Nagatsuta, Midori-ku, Yokohama 226-8503, Japan}
\date{\today}
\begin{abstract}
We have studied the electronic structure 
of epitaxially grown thin films of La$_{1-x}$Sr$_x$FeO$_3$ by 
{\it in-situ} photoemission 
spectroscopy (PES) and x-ray absorption spectroscopy (XAS) 
measurements. 
The Fe 2$p$ and valence-band PES spectra and the O $1s$ XAS 
spectra of LaFeO$_3$ have been successfully reproduced 
by configuration-interaction cluster-model 
calculation and, except for the satellite structure, 
by band-structure calculation.
From the shift of the binding energies of core levels, 
the chemical potential was found to be shifted downward 
as $x$ was increased. 
Among the three peaks in the valence-band spectra of
 La$_{1-x}$Sr$_x$FeO$_3$, 
the peak nearest to the Fermi level ($E_F$), due to the ``$e_{g}$ band'', 
was found to move toward $E_F$ and became weaker as $x$ was increased, whereas 
the intensity of the peak just above $E_F$ in the O $1s$ XAS spectra 
increased with $x$. 
The gap or pseudogap at $E_F$ was seen for all values of $x$. 
These results indicate that changes in the spectral line shape around $E_F$ 
are dominated by spectral weight transfer from below to above $E_F$ 
across the gap and are therefore highly 
non-rigid-band-like.
\end{abstract}
\maketitle
\section{Introduction}
Since the discovery of high-$T_c$ superconductivity in the cuprates, 
great interest has revived in 
perovskite-type transition-metal oxides because of  
their intriguing properties, such as metal-insulator transition (MIT), 
colossal magnetoresistance (CMR), and ordering of spin, charge, and
orbitals \cite{rev}.
In most cases, hole doping plays a crucial role in realizing these
interesting physical properties. However, there has been little
consensus on how the electronic structure evolves with hole doping in 
these compounds, namely, whether new states are created in the gap or 
the chemical potential is simply shifted as in the rigid-band model, 
even in the most extensively studied case 
of the high-$T_c$ cuprates \cite{vee,eisaki,chen}. 
Among the perovskite-type transition-metal oxides,  
La$_{1-x}$Sr$_x$FeO$_3$ (LSFO) has attracted much interest because it 
undergoes a pronounced charge disproportionation and an associated MIT 
around $x\simeq 2/3$ \cite{Takano}. 
One end member, LaFeO$_{3}$, is an antiferromagnetic insulator
with a high N\'eel temperature ($T_N=738$ K). 
The character of the band gap is of the charge-transfer (CT) type, and 
the optical gap is $\sim$ 2.1 eV \cite{opt1}. 
The other end member, SrFeO$_{3}$, is a helical antiferromagnetic metal
with $T_N=134$ K. 
In an early photoemission study, it was found that 
its ground state is dominated by the $d^5\underline{L}$ configuration, 
where $\underline{L}$ denotes a hole in the O $2p$ band, rather 
than the $d^4$ configuration \cite{Bocquet}, meaning that the system
is a negative-charge-transfer-energy compound 
and that holes in the oxygen $2p$ band are responsible for 
the metallic behavior \cite{Mizokawa,Kawakami} . 
One striking feature of LSFO is that the insulating phase is unusually 
wide in the phase diagram 
(especially at low temperatures $0<x<0.9$, and even at room temperature 
$0<x<0.5$) \cite{Matsuno}. The O 1$s$ x-ray absorption spectroscopy
(XAS) study by Abbate {\it et al.} \cite{Abbate} has suggested 
that doped holes have the O 2$p$ character. 
Therefore, the central question for this system remains unresolved of 
how does the electronic 
structure evolve from a CT insulator LaFeO$_3$ 
to the metallic oxygen holes of SrFeO$_3$ as a function of hole doping. 

Recently, high-quality perovskite-type oxide single-crystal thin films 
grown by pulsed laser deposition (PLD) 
have become available \cite{Izumi,Choi}, 
and a setup has been developed for 
their {\it in-situ} photoemission measurement \cite{Kumi,Horiba}. 
In the present work, we address the above questions about the electronic 
structure of LSFO 
by measuring soft x-ray photoemission and absorption spectra of 
epitaxially-grown high-quality thin films prepared {\it in situ}. 
A systematic x-ray photoemission study of scraped 
bulk LSFO samples has been reported by Chainani {\it et al.} \cite{Chainani}. 
Structures in the valence band, however, were not clearly resolved 
partly because of 
the limited energy resolution ($\sim 0.8$ eV).
In a more recent study by Matsuno {\it et al.} using similarly 
prepared samples \cite{Matsuno}, 
detailed temperature-dependent 
changes near the Fermi level ($E_F$) were studied with high energy 
resolution. However, the O $2p$ cross section 
overwhelmed the Fe $3d$ emission for the low photon energies $20\alt
h\nu\alt 100$ eV used in their study \cite{table}. 

In the present study, we have used soft x-rays 
with high-energy resolution ($\sim 200$ meV) and succeeded in resolving detailed 
spectral features and in directly obtaining more information 
about the Fe $3d$ states.
We could determine the Fe $3d$ contribution more clearly with better 
bulk sensitivity due to the longer photoelectron escape depth for 
the higher photoelectron kinetic energies \cite{bulk}. 
We emphasize that in the present work the usefulness of bulk 
sensitivity is further enhanced by the use of high-quality 
epitaxial thin film samples \cite{Kumi}. 
Combining the soft x-ray photoemission spectra and 
O $1s$ XAS spectra, we have successfully obtained a picture of 
how the electronic structure evolves from a CT insulator to 
an oxygen-hole metal through the observation of the chemical potential shift and 
spectral weight transfer.
\section{Experiment}
\begin{figure}
\begin{center}
\includegraphics[width=5cm]{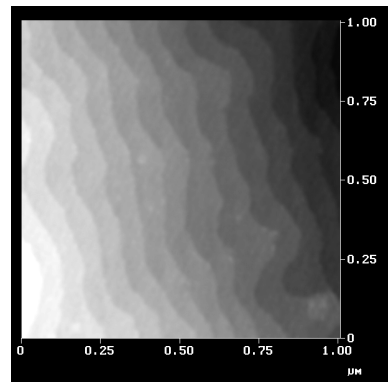}
\caption{AFM image of an epitaxially grown LaFeO$_3$ thin film. 
Image size is 1 $\mu$m $\times$ 1 $\mu$m.}
\label{AFM}
\end{center}
\end{figure}

The photoemission spectroscopy (PES) and XAS measurements 
were performed at BL-2C of Photon Factory, High Energy Accelerators 
Research Organization (KEK), using a combined 
Laser molecular beam epitaxy (MBE) and photoemission 
spectrometer system. 
Details of the experimental setup are described in Ref \cite{Horiba}. 
Epitaxial films of LSFO were fabricated 
by the PLD method. 
Single crystals of Nb-doped SrTiO$_3$ were used as substrates. 
Nb-doping was necessary to avoid charging effects during the PES 
measurements. 
A Nd:YAG laser was used for ablation in its frequency-tripled mode 
($\lambda=355$ nm) at a repetition rate of 0.33 Hz. 
The substrates were annealed at 1050${}^{\circ}$C at an oxygen pressure of 
$\sim 1\times 10^{-6}$ Torr to obtain an atomically flat TiO$_2$-terminated 
surface \cite{kawasaki}. 
LSFO thin films of $\sim$ 100 monolayers were deposited on the
substrates at 
950${}^{\circ}$C at an oxygen pressure of 
$\sim 1\times 10^{-4}$ Torr. The films were post-annealed at 400${}^{\circ}$C 
at an atmospheric pressure of oxygen to remove oxygen vacancies.
The samples were then transferred from the MBE chamber to the spectrometer 
under ultrahigh vacuum. 
The surface morphology of the measured films was checked 
by {\it ex-situ} atomic force microscopy (AFM). 
The AFM image of a LaFeO$_3$ thin film in Fig.~\ref{AFM} 
shows atomically flat step-and-terrace structures. 
The crystal structure was characterized by four-circle X-ray diffraction, 
and coherent growth on the substrate was observed.
All the photoemission measurements were performed under an ultrahigh vacuum 
of $\sim 10^{-10}$ Torr at room temperature.
The PES spectra were measured using 
a Scienta SES-100 electron-energy analyzer. 
The total energy resolution was about 200-500 meV depending on photon
energy. 
The Fermi level ($E_F$) position was determined by measuring gold 
spectra. 
The XAS spectra were measured using the total-electron-yield method.   

The stoichiometry of thin films was carefully characterized 
by analyzing the relative intensity of the relevant core levels, 
confirming that the composition of samples is almost the same as ceramic
targets. 
\section{Results and discussion}
\subsection{Electrical resistivities}
\begin{figure}
\begin{center}
\includegraphics[width=7cm]{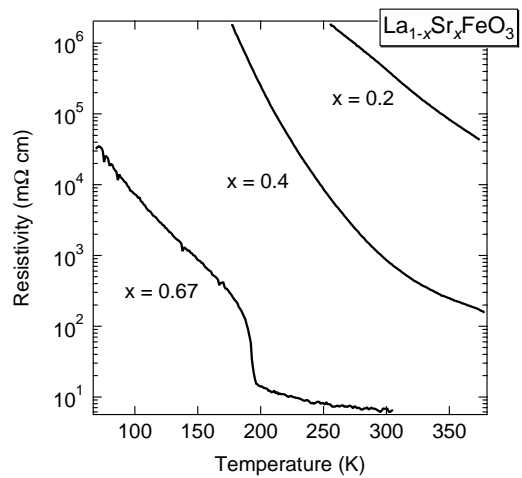}
\caption{Electrical resistivity of La$_{1-x}$Sr$_x$FeO$_3$ thin films.}
\label{pt}
\end{center}
\end{figure}

Figure \ref{pt} shows the electrical resistivity of 
La$_{1-x}$Sr$_x$FeO$_3$ thin films samples 
which were made under the same condition as the 
samples for the photoemission measurements. 
The samples for the resistivity measurements 
were grown on Nb-free SrTiO$_3$ substrates 
to prevent the electric current from flowing through 
the conducting substrate. 
As for $x=0.67$, there is a jump of resistivity, caused by charge
disproportionation, almost at the 
same temperature as the bulk 
samples \cite{Matsuno,opt2} ($T_{CD}=190$ K). 
\footnote{This transition is often called ``metal-insulator
transition'', but the electrical resistivity satisfies $d\rho/dT < 0$ 
even above the transition temperature.} 
\subsection{LaFeO$_3$}
\begin{figure}
\begin{center}
\includegraphics[width=6cm]{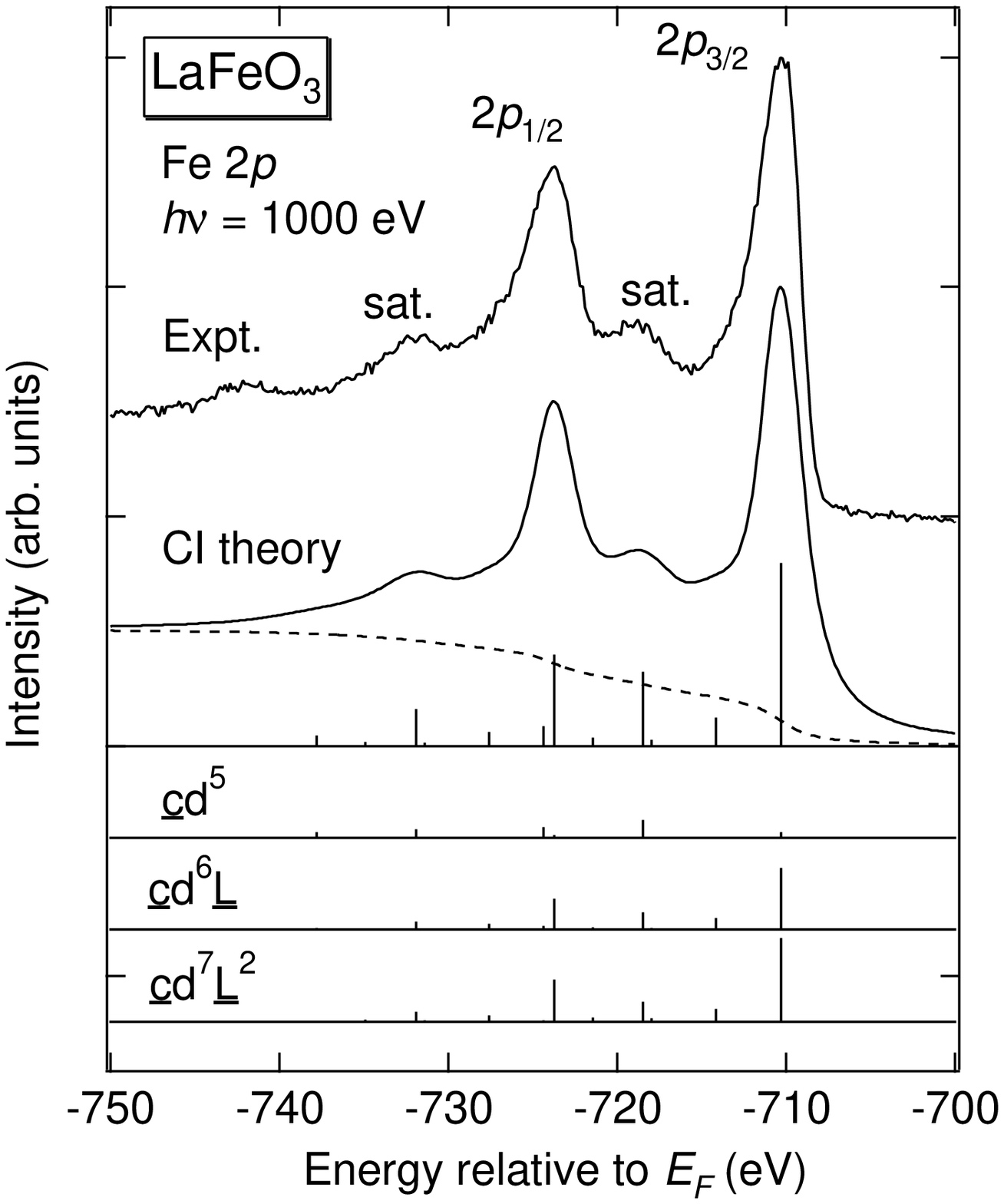}
\includegraphics[width=6cm]{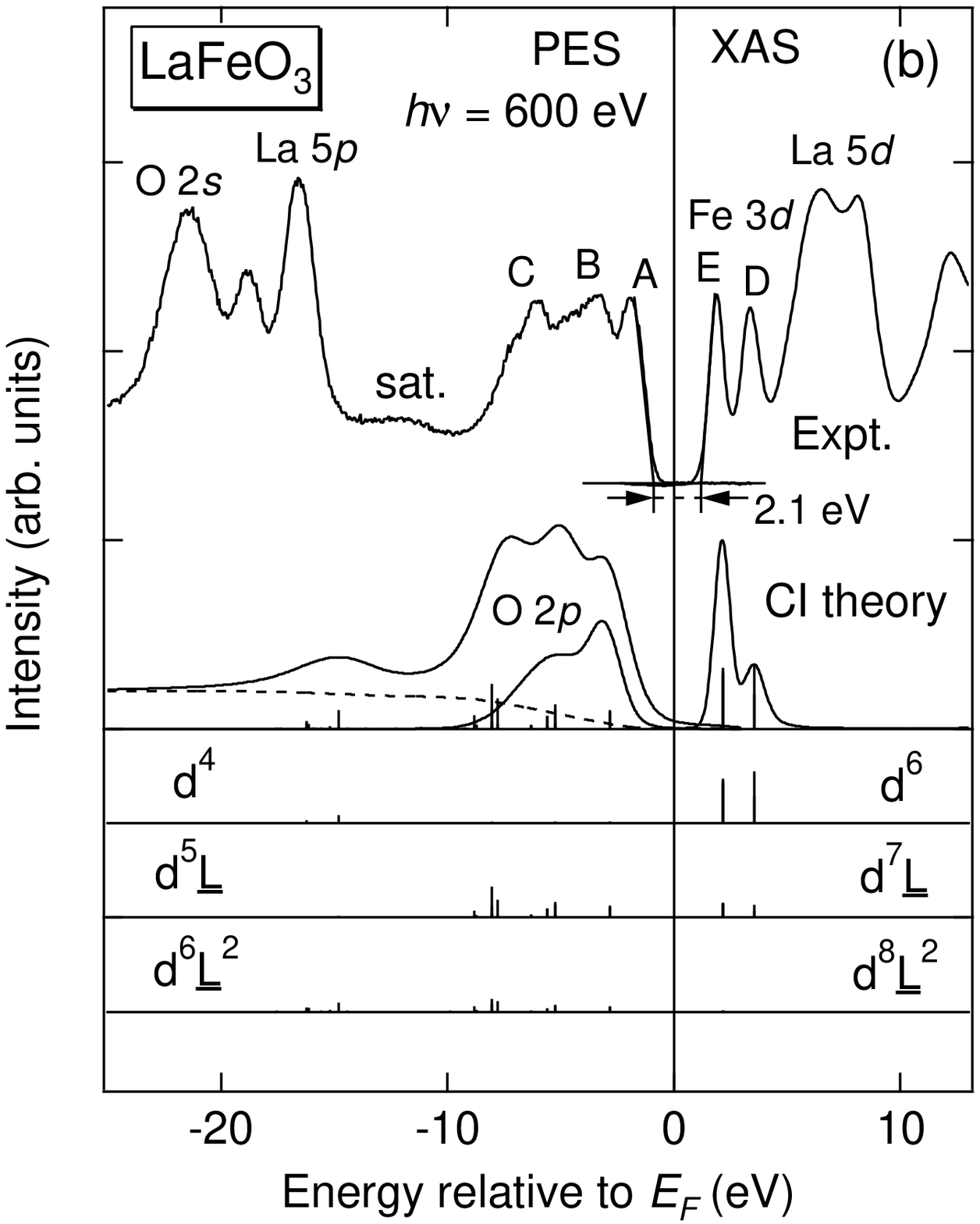}
\caption{Photoemission and XAS spectra of LaFeO$_3$ 
and their CI cluster-model analyses. 
In the bottom panels, final states are decomposed into 
contributions from different configurations. 
The dotted lines indicate the background due to secondary electrons. 
(a) Fe $2p$ core-level photoemission spectrum.  
(b) Valence-band photoemission and O $1s$ XAS spectra.}
\label{LFO}
\end{center}
\end{figure}

Before proceeding to the composition dependence of LSFO, we first 
characterize the electronic structure of the La-end composition, LaFeO$_3$ (LFO). 
Figure \ref{LFO} gives the Fe $2p$ PES spectrum, the valence-band 
PES spectrum, and the O $1s$ XAS spectrum of LFO.
As the $E_F$ position for O $1s$ XAS cannot be determined 
unambiguously from the O $1s$ core-level PES and XAS data 
because of the unknown effect of the core-hole potential \cite{Saito}, 
the XAS spectrum has been aligned so that 
the gap magnitude agrees with that obtained from the optical
measurements, 2.1 eV \cite{opt1}, as shown in Fig.~\ref{LFO} (b). 
In the main valence-band region [$-(0-10)$ eV], 
we observe three structures denoted by A, B, and C 
and a satellite structure at $\sim -12$ eV. 
Structure A is assigned to the $e_g$ states of Fe $3d$, structure B to 
the $t_{2g}$ states, and structure C to the Fe 3$d$ $-$ O 2$p$ 
bonding states. 
The O $1s$ XAS spectrum shows empty Fe 3$d$ states 
within $\sim$ 5 eV of $E_F$ and the La $5d$ states above it. 
The empty Fe $3d$ states are 
split into two peaks, D and E, 
due to the $e_g-t_{2g}$ crystal-field splitting. 

In order to interpret those spectra quantitatively, 
we performed a configuration-interaction (CI) cluster-model 
calculation \cite{cluster1,Bocquet}. 
We considered an [FeO$_6$]$^{9-}$ octahedral cluster. 
In this model, the ground state is described as
\begin{equation}
\Psi_g=\alpha_1|d^5\rangle+\alpha_2|d^6\underline{L}\rangle+
\alpha_3|d^7\underline{L}^2\rangle+\cdots.
\end{equation} 
Our model, in which $d$ electrons are 
assumed to be localized in the cluster, is considered to be a good
approximation since LaFeO$_3$ is an insulator and 
the LDA$+U$ calculation has shown that the Fe $3d$ states in LaFeO$_3$ have
weak band dispersion, so the translational symmetry does not affect the
angle-integrated photoemission spectra of the Fe 3d band significantly. 
The final state for the emission of an Fe $2p$ core electron is given by 
\begin{equation}
\Psi_f=\beta_1|\underline{c}d^5\rangle+\beta_2|\underline{c}d^6\underline{L}\rangle+
\beta_3|\underline{c}d^7\underline{L}^2\rangle+\cdots,
\end{equation}
where $\underline{c}$ denotes an Fe $2p$ core hole. 
The final state for the emission of an Fe $3d$ electron is given by
\begin{equation}
\Psi_f=\gamma_1|d^4\rangle+\gamma_2|d^5\underline{L}\rangle+
\gamma_3|d^6\underline{L}^2\rangle+\cdots,
\end{equation}
and that for O $1s$ XAS by
\begin{equation}
\Psi_f=\delta_1|d^6\rangle+\delta_2|d^7\underline{L}\rangle+
\delta_3|d^8\underline{L}^2\rangle+\cdots.
\end{equation}
The O $1s$ XAS spectrum represents the unoccupied O $2p$ partial 
DOS, and since other orbitals are strongly hybridized with the 
O $2p$ orbitals, 
O $1s$ XAS also reflects the empty Fe $3d$ and La $5d$ bands. 

To calculate the valence-band PES spectrum, 
the O $2p$ emission spectrum has to be added to the Fe $3d$ 
spectrum. 
The line shape of the O $2p$ band was taken from 
the PES data of La$_{0.33}$Sr$_{0.67}$FeO$_3$, measured at 
$h\nu=21.2$ eV \cite{thesis}. 
In order to take into account the chemical potential shift 
between LaFeO$_3$ and La$_{0.33}$Sr$_{0.67}$FeO$_3$ (see Fig.~\ref{shift1}), 
the $h\nu=21.2$ eV spectrum has been shifted downward by 0.78 eV. 
The relative intensities of the Fe 3$d$ - and O 2$p$ - derived 
spectra have been determined from the atomic photoionization 
cross sections \cite{table} with the O 2$p$ intensity multiplied 
by a factor of $\sim$ 3 \cite{Sawatzky}. 
Parameters to be fitted are the charge-transfer energy 
from the O $2p$ orbitals to the empty Fe $3d$ orbitals denoted by $\Delta$, 
the $3d-3d$ on-site Coulomb interaction energy denoted by $U$, and the
hybridization strength between the Fe $3d$ and O $2p$ orbitals denoted by 
Slater-Koster parameters $(pd\sigma)$ and $(pd\pi)$. 
The ratio $(pd\sigma)/(pd\pi)=-2.2$ has been assumed, as usual 
\cite{cluster1,Bocquet}. 
The configuration dependence of the transfer integrals 
has been taken into account \cite{DMS1}. 
Racah parameters are fixed at the 
free ion values of Fe$^{3+}$ ($B=0.126$ eV, $C=0.595$ eV) 
\cite{Tanabe}.  
We took into account the intra-atomic 
multiplet coupling for the valence-band spectrum, 
whereas it was not taken fully into account for the Fe $2p$ 
spectrum as in the case of \cite{Bocquet}.  

The calculated Fe $2p$ core-level photoemission 
spectrum has been broadened 
with an energy-dependent Lorentzian with FWHM
\begin{equation}
 2\Gamma = 2\Gamma_0(1+\alpha \Delta E),
\end{equation} 
where $\Delta E$ denotes the energy separation from the main peak. 
We adopted the values $\alpha=0.15$ and $\Gamma_0=1.2$ eV. 
We then used a Gaussian broadening of 1.0 eV to simulate the 
instrumental resolution and broadening due to  
the core hole-$3d$ multiplet coupling. 
The calculated valence-band spectrum has been broadened with a 
Gaussian of 1.6 eV FWHM and an energy-dependent 
Lorentzian ($\mbox{FWHM}=0.2|E-E_F|$ eV) \cite{Hufner} 
to account for the combined effects of the 
instrumental resolution and the $d$ band dispersion, 
and the lifetime 
broadening of the photohole, respectively. 

The best-fit results have been obtained 
setting $\Delta =2.0$ eV, $U=6.0$ eV, and $(pd\sigma)=-1.9$
eV as shown in Fig.~\ref{LFO}. 
The $e_{g}-t_{2g}$ crystal-field parameter of $10Dq=0.41$ eV was 
assumed to 
reproduce the splitting in the O $1s$ XAS spectrum. 
These parameters are consistent with previously reported ones, 
which showed that LaFeO$_3$ is a charge-transfer-type insulator, where $\Delta < U$
\cite{Bocquet}. However, the value of $U$ had to be taken 
smaller than the previously
reported value ($U=7.5$ eV) \cite{Bocquet} in order to reproduce 
both the Fe $2p$ core level and valence-band spectra simultaneously. 
As we have succeeded in reproducing both the core level and the valence band 
using the same parameter set, 
the present results are more accurate than previous ones
\cite{Bocquet}, although the PES spectrum calculated with this $U$ 
value is still slightly too deep and the band gap is  
overestimated as shown in Fig.~\ref{LFO} (b). 
Also, there is a significant difference in the calculated and 
the experimental satellite energy positions. 
For the Fe $2p$ spectrum, we conclude that the main peaks 
mostly come from $\underline{c}d^6\underline{L}$ and 
$\underline{c}d^7\underline{L}^2$ final states, 
while there is significant contribution of 
the $\underline{c}d^5$ final states to the satellite. 
Good agreement is obtained between the calculated and 
the experimental satellite energy position, in contrast to 
the valance band. As mentioned above, the effect of 
the intra-atomic multiplet is treated differently in the core-level and
valence-band CI calculations, 
and this point probably causes this difference. 
Therefore, the precise agreement of the satellite position of the
core-level spectrum may be rather fortuitous. 
For the valence band, we conclude that the three main structures
are derived from $d^5\underline{L}$ and $d^6\underline{L}^2$ final
states with admixture of the O $2p$ band, 
while the satellite has strong contribution from 
the $d^4$ final state. 
Due to the small value of $\Delta$, 
final states with two holes at the ligand site are 
important for the interpretation of the PES spectra. 
As for the O $1s$ XAS spectrum, the final states have mostly $d^6$ 
character, and
therefore one can interpret the data without significant 
contributions from charge-transfer 
($d^7\underline{L},d^8\underline{L}^2,\cdots$) final states.      

\begin{figure}
\begin{center}
\includegraphics[width=8cm]{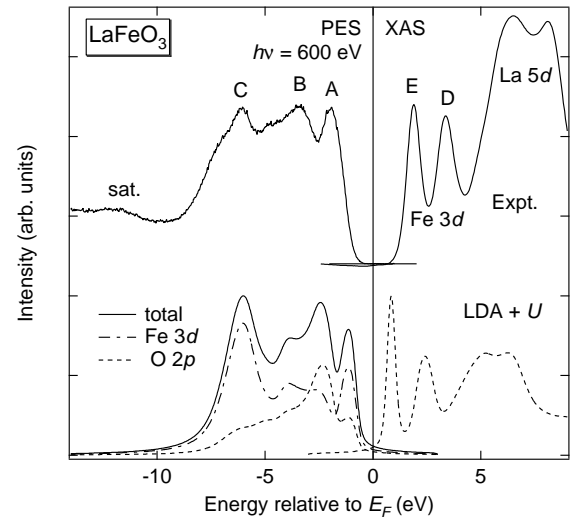}
\caption{Comparison of the photoemission and O 1$s$ XAS spectra of 
LaFeO$_3$ with the 
LDA $+U$ band-structure calculation.}
\label{LDA}
\end{center}
\end{figure}

We have also compared the spectra with the LDA $+U$ band-structure calculation 
in Fig.~\ref{LDA}. 
The full-potential linearized augmented-plane-wave (FLAPW) method 
was employed
with an exchange-correlation potential of Vosko, Wilk and
Nusair \cite{LDA1}. 
The effective Coulomb interaction parameter, $U_{eff}\equiv U-J$, 
in the LDA $+U$ scheme
was set to 2 eV for all the Fe 3$d$ orbitals \cite{LDA2}. 
A G-type antiferromagnetic state was assumed. 
The calculated DOS has been broadened 
with a Gaussian of 0.15 eV FWHM and an energy-dependent 
Lorentzian ($\mbox{FWHM}=0.2|E-E_F|$ eV) \cite{Hufner} 
to account for the instrumental resolution and the lifetime 
broadening of the photohole, respectively. 
Below $E_F$, we have added the partial DOS of O $p$ and Fe $d$, considering
their cross sections at $h\nu=600$ eV \cite{table}, with the 
multiplication factor of $\sim$ 3 for the O $p$ 
partial DOS \cite{Sawatzky}. Above $E_F$, we have
considered only the partial DOS of O $p$ because the O 1$s$ XAS 
spectrum is due to the dipole-allowed transition from the 
O 1$s$ core level. 
The three main structures, A, B, and C, were successfully reproduced 
(including a weak shoulder between B and C), 
consistent with a previous report \cite{Sarma}. 
However, the calculated band gap of 1.3 eV was too
small compared with the optical gap of 2.1 eV \cite{opt1}, and the satellite
structure could not be reproduced since our value of $U_{eff}=2$ eV was
chosen 
to be the best value for the reproduction of the main structures, 
not of the satellite structure and the value of the band gap. 

From this comparison, we conclude that the valence-band spectra of LFO 
can be explained well both by the CI cluster-model calculation and, 
except for the satellite 
structure, by the LDA $+U$ band-structure calculation. 
It is a reasonable result that both calculations can reproduce 
the Fe $3d$ - band region of LaFeO$_3$ equally well because, 
except for the satellite, there is one-to-one correspondence between 
the two calculations for the peak positions and the orbital 
character for the $e_g$ band (structure-A), $t_{2g}$ band (structure-B), 
and Fe 3$d$ $-$ O 2$p$ bonding states (structure-C).
\subsection{La$_{1-x}$Sr$_x$FeO$_3$}
\begin{figure}
\begin{center}
\includegraphics[width=8.5cm]{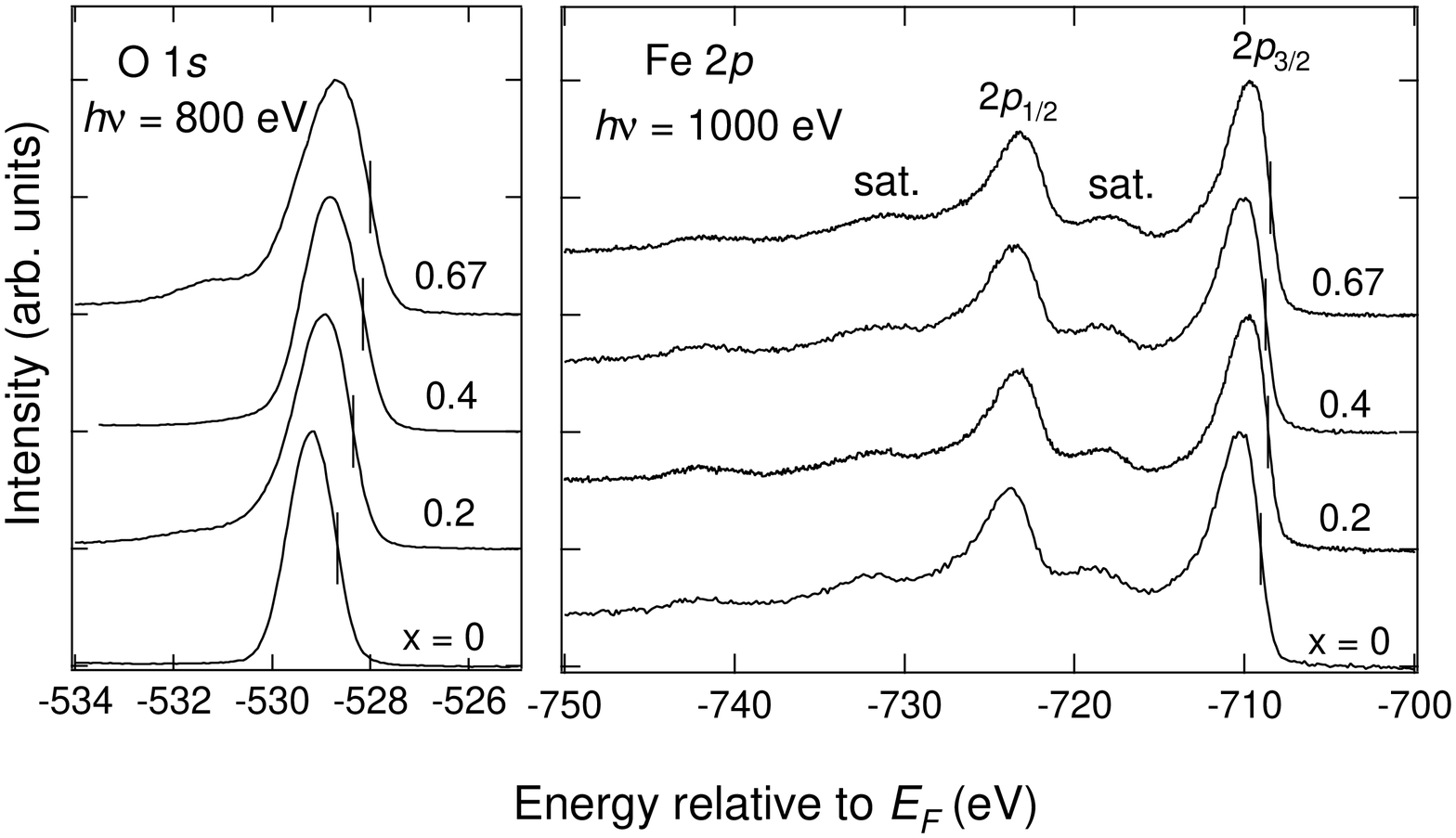}
\includegraphics[width=8.5cm]{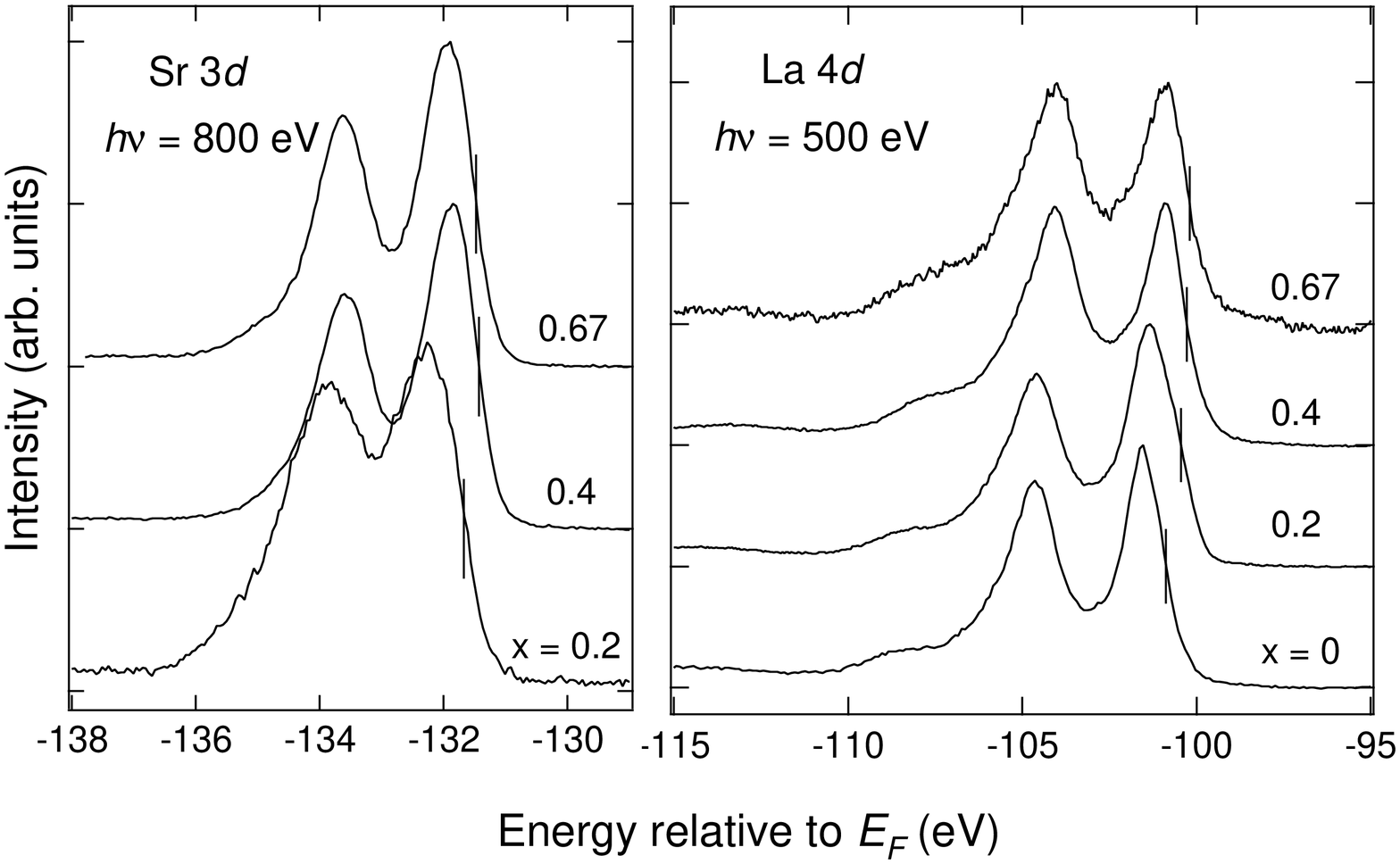}
\caption{Core-level photoemission spectra of La$_{1-x}$Sr$_x$FeO$_3$.}
\label{core}
\end{center}
\end{figure}

Next, we go to the question of how the electronic 
structure of LFO changes upon hole doping. 
Figure \ref{core} shows the core-level photoemission spectra 
of La$_{1-x}$Sr$_x$FeO$_3$. 
The ``contamination'' signal 
on the higher binding energy side of the O $1s$ peak was weak 
enough, except for the $x=0.67$ sample, 
indicating that the surface was reasonably clean. 
The line shape of the Fe $2p$ core level has almost no composition
dependence, consistent with the picture that doped holes go into 
states of primarily O $2p$ character, and not of Fe $3d$ character \cite{Abbate}. 

\begin{figure}
\begin{center}
\includegraphics[width=9cm]{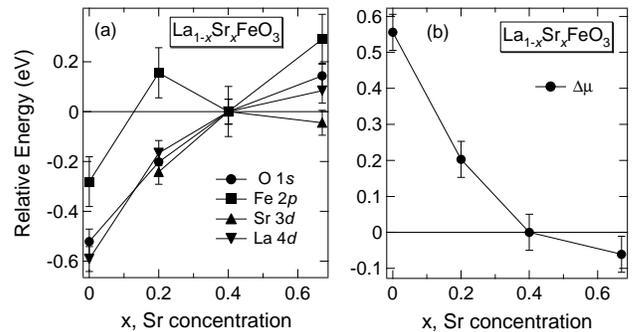}
\caption{Binding energy shifts of spectral features in
 La$_{1-x}$Sr$_x$FeO$_3$. (a) Core levels. (b) Chemical potential shift 
deduced from the O $1s$, Sr $3d$, and La $4d$ core levels.}
\label{shift1}
\end{center}
\end{figure}

All the core-level spectra, except for the Fe $2p$ core level, are 
shifted toward lower 
binding energies with $x$, as plotted in Fig.~\ref{shift1} (a). 
Here, the midpoint of the lower binding energy slope 
is taken as representing the shift of the peaks because 
this part is generally 
least affected by possible contamination \cite{Ino,Matsuno2}.
In determining the ``Relative energy'', we adopted the sample of $x=0.4$ 
as the zero. The sample of $x=0$ has no Sr, 
so we cannot adopt this as the zero. 
We consider that it is reasonable to adopt $x=0.4$ as the zero 
since $x=0.4$ is almost in the mid-position of this hole-doping system although it is not the unique choice. 

\begin{figure}[t]
\begin{center}
\includegraphics[width=8cm]{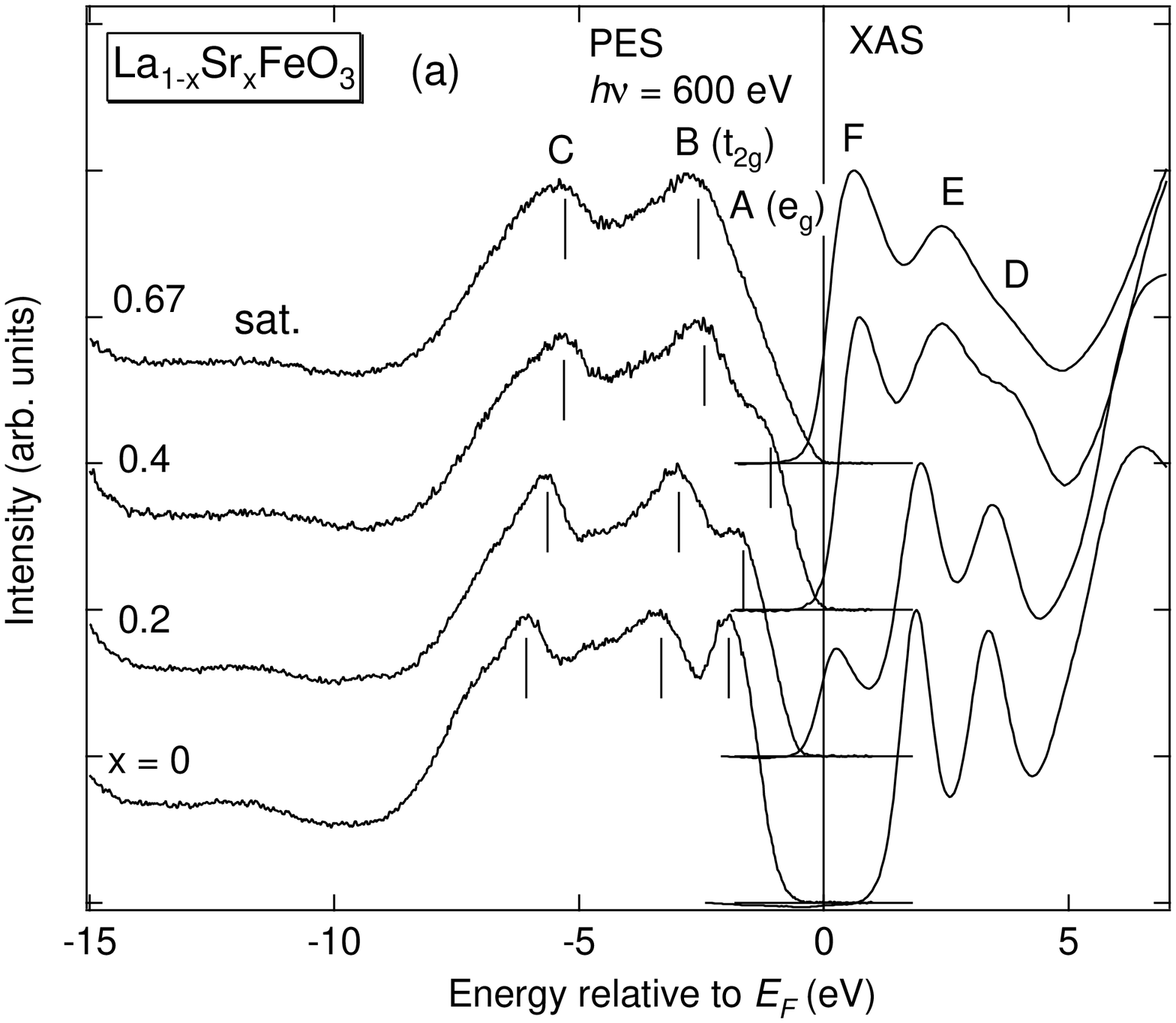}
\includegraphics[width=5cm]{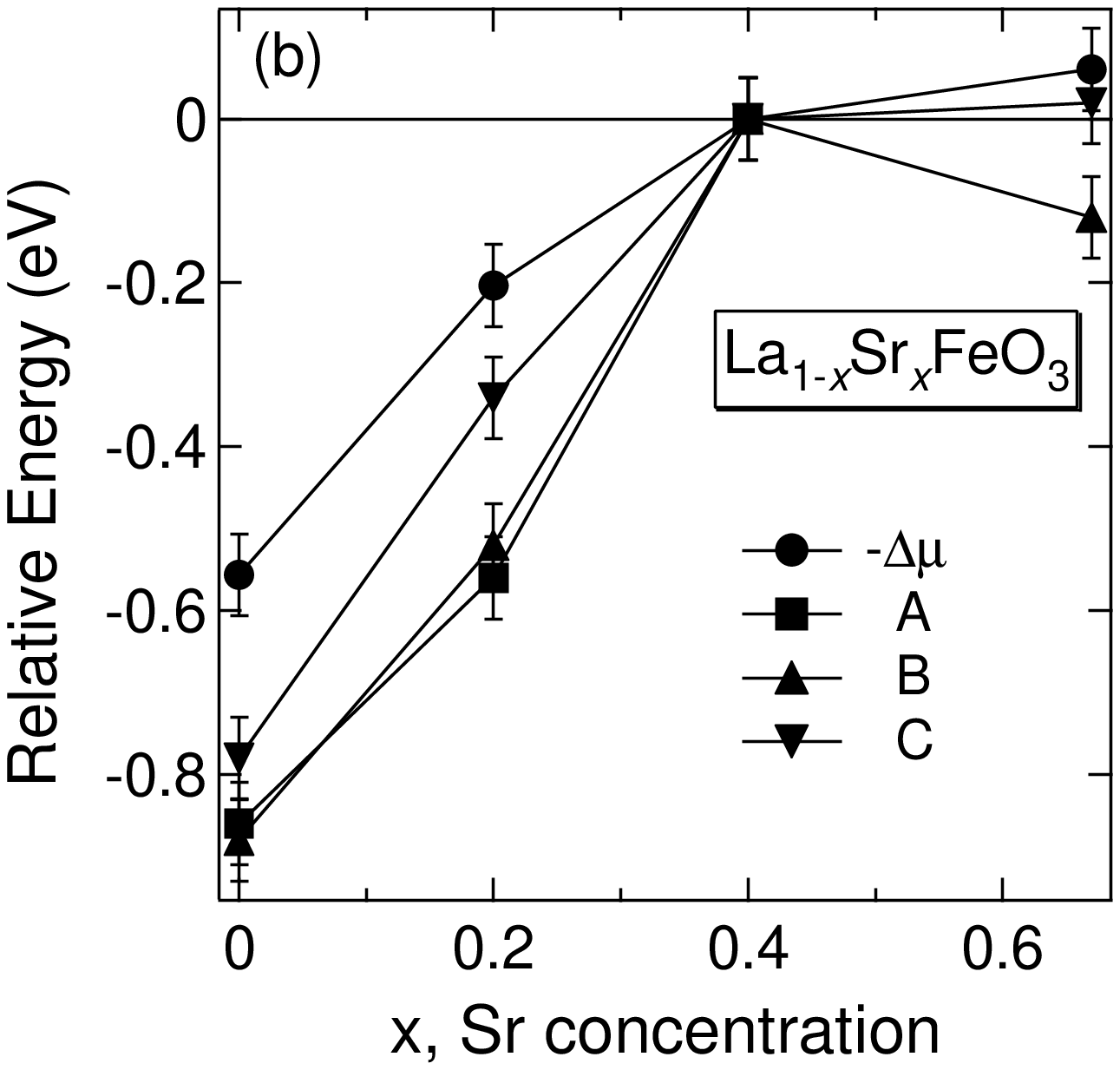}
\includegraphics[width=8.5cm]{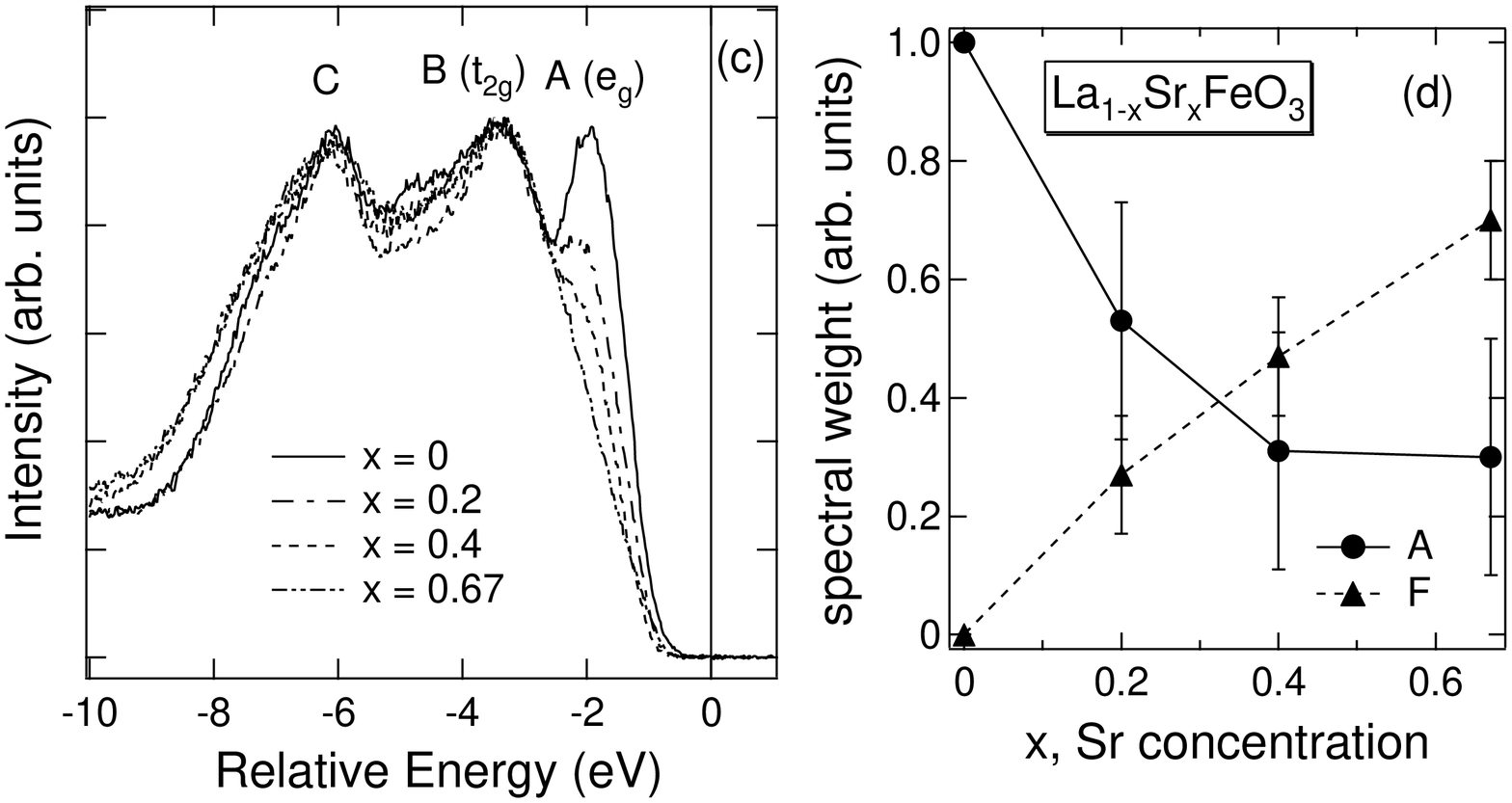}
\caption{PES and XAS spectra of La$_{1-x}$Sr$_x$FeO$_3$. 
(a) Combined PES and XAS spectra. 
(b) Shifts of structures in the valence band. 
(c) Valence-band photoemission spectra shifted so that 
structures A, B, and C are aligned. 
(d) Spectral weight of structures A and F as functions of $x$.}
\label{val1}
\end{center}
\end{figure}

The shift $\Delta E_B$ of a core-level binding energy measured from 
$\mu$ is given by 
\begin{equation}
 \Delta E_B=\Delta\mu+K\Delta Q+\Delta V_M+\Delta E_R ,
\end{equation}
where 
$\Delta\mu$ is the change in the chemical potential, 
$\Delta Q$ is the change in the number of valence electrons on the atom
considered,
$\Delta V_M$ is the change in the Madelung potential, and 
$\Delta E_R$ is the change in the core-hole screening \cite{Hufner-book}.
As seen in Fig.~\ref{shift1} (a) the Fe $2p$ core level moves in a different way 
from the other core levels probably 
because the formal valence of Fe changes with 
hole doping, reflecting both the chemical potential
shift and the 
``chemical shift'', which is due to the increase in the Fe 
valence with hole doping ($\propto -\Delta Q$), 
as in other transition-metal oxides \cite{Ino,Matsuno2,pote}. 
The similar shifts of the O $1s$, Sr $3d$, and La $4d$ core levels indicate 
that the change in the Madelung potential ($\Delta V_M$) is negligibly
small because it would shift the core levels of anions and cations in 
different directions.  
Core-hole screening by conduction electrons is also considered to 
be negligibly small in transition-metal oxides \cite{Ino,Matsuno2,pote}.
Therefore, we take the average of the shifts of these three core levels 
as a measure of $\Delta\mu$ in LSFO. 
Figure \ref{shift1} (b) shows $\Delta\mu$ thus determined plotted 
as a function of $x$. 
The shifts become slightly weaker above $x=0.4$. 
In the rigid band picture, $-{\partial\mu}/{\partial x}$ is inversely 
proportional to the DOS at $E_F$ \cite{pote}, which may explain the
weakening of the chemical potential shift. 
However, there is 
a gap (absence of finite DOS at $E_F$) or 
a pseudogap (depression of DOS at $E_F$, see Fig.~\ref{val1} (a)) 
for all the compositions, 
which means that interpretation beyond the rigid band picture 
is necessary. 
La$_{2-x}$Sr$_x$CuO$_4$ is a typical example in which 
the suppression of the chemical potential shift has been observed 
in the underdoped region, where there is a pseudogap at $E_F$ \cite{Ino}. 
This phenomenon has been attributed to the formation of 
charge stripes, a kind of ``microscopic phase separation''
in which the distance between hole stripes decreases with hole concentration $x$. 
Further studies up to $x=1$ are necessary to see whether 
the weakening of the shifts at large $x$'s is related to the charge
disproportionation or not. 

Figure \ref{val1} (a) shows the doping dependence of  
the combined valence-band photoemission 
and the O $1s$ XAS spectra. 
Here, the Fermi levels of the XAS spectra for various $x$ have 
been determined by combining the Fermi level position in the 
LFO spectrum with the $x$-dependent shift of the O $1s$ 
core-level peak for the sake of convenience 
\footnote{It has been recognized for many years 
that XAS spectra, in particular O $1s$ XAS in transition-metal oxides, 
do not precisely represent unoccupied DOS and also that it 
is difficult to determine the exact $E_F$ position. For example, the BIS 
spectra (K. Morikawa {\it et al.}, Phys. Rev. B 52, 13711 (1995)) and the O $1s$ 
XAS spectra (H. I. Inoue {\it et al.}, Physica C 235-240 1007 (1994)) of CaVO$_3$ and 
SrVO$_3$ are different. The O $1s$ XAS spectra show a build-up of 
spectral weight at the leading edge, which is not present in the real 
DOS measured by BIS. Therefore, the tail of the XAS spectra of the doped
samples does not necessarily mean that the DOS is finite at $E_F$.}. 
In the PES spectra, one can observe the three 
main structures A ($e_g$ band), B ($t_{2g}$ band), and C (Fe 3$d$ $-$ O
2$p$ bonding states) 
and the satellite, as in the case 
of LFO. A gap (absence of finite DOS at $E_F$) 
or a pseudogap (depression of DOS at $E_F$) 
was seen for all values of $x$, 
as was observed for La$_{1-x}$Sr$_x$MnO$_3$ \cite{Saito}.
The sample of $x=0.67$ undergoes a metal-insulator transition at 190 K, 
but there is little DOS at $E_F$ at room temperature. 
This may be related to the fact 
that the electrical resistivity satisfies $d\rho/dT < 0$ 
even above the transition temperature as shown in Fig.~\ref{pt}. 
The intensity of the satellite has almost no composition dependence.  
Figure \ref{val1} (b) shows the binding energy shifts of structures A, B,
and C plotted against $x$. 
Structures A-C move toward $E_F$ upon hole doping up to $x=0.4$. 
These shifts are in good agreement 
with the core level shifts, indicating the rigid-band shift 
occurs in the valence band. 
In addition, structure A, which is assigned to the ``$e_{g}$ band'', 
becomes weaker 
with increasing $x$, indicating 
that holes are doped into the ``$e_{g}$ band'', and is finally
obscured at $x=0.67$. 
The weakening of structure A with $x$ is more clearly seen in 
Fig.~\ref{val1} (c), where the energy positions of structures B and C have been aligned. 
In the XAS spectra, a new peak F grows within the band 
gap of LFO upon hole doping, as seen in the previous 
study \cite{Abbate}. 
Since $x=0$, 0.2, 0.4 samples are insulating, 
structure A and F are separated by a gap. 
Therefore, this structure F cannot be part of structure A 
but a state created in the gap. 
The combined PES and XAS spectra thus demonstrate that spectral weight 
is transferred from structure A below $E_F$ to structure F above $E_F$, and 
the band gap is filled by the new spectral weight F as holes are doped. 
Spectral weight of structures A and F is plotted as a function of Sr
concentration $x$ in 
Fig.~\ref{val1} (d). The weight of structure F is almost proportional to $x$, 
indicating that doped holes go into this structure. 
Also, $E_F$ is located within the gap or the pseudogap for all
$x$'s, that is, 
the intensity at $E_F$ remains always small (more clearly in the 
PES spectrum), which may correspond to the wide insulating region in the 
LSFO phase diagram. 
This non-rigid-band
behavior within $\sim 2$ eV of $E_F$ is apparently in conflict 
with the monotonic chemical potential shift. 
We therefore conclude that in this system the effect of hole-doping 
can be described in the framework of the rigid-band model as far as 
the shifts of the spectral features are concerned, whereas the ``$e_{g}$
band'' 
shows highly non-rigid-band-like 
behavior with transfer of spectral weight from below $E_F$ to above it 
across the gap or pseudogap at $E_F$.
\section{Conclusion}
We have studied the composition-dependent electronic structures 
of LSFO using epitaxial thin films by {\it in-situ} PES and XAS 
measurements. By using soft x-rays of high energy resolution and high-quality
sample surfaces, we succeeded in obtaining high-quality spectra with detailed 
spectral features of high bulk sensitivity. 
The Fe 2$p$ and valence-band PES spectra and the O $1s$ XAS 
spectra of LaFeO$_3$ have been successfully reproduced 
by CI cluster-model calculation as well as by band-structure 
calculation. This is considered to be a natural result because 
there is a one-to-one correspondence between both calculations 
for the structures in the valence band. 
The shift of the chemical potential was found to become slightly weak 
above $x=0.4$. 
Further studies up to $x=1$ are necessary to see whether 
this weakening is related to the charge disproportionation around $x=0.67$. 
In the valence-band spectra, the structure nearest to $E_F$ becomes 
weaker and moves toward $E_F$ as $x$ is increased. 
The gap or pseudogap at $E_F$ was seen for all compositions. 
These results mean that the simple rigid band model does not work in
this system and the transfer of spectral weight 
occurs across $E_F$ in a highly non-rigid-band-like manner.
\section{Acknowledgment}
The authors would like to thank A. Tanaka for informative discussions. 
We are also grateful to K. Ono and A. Yagishita for their support 
at KEK-PF. 
This work was done under the approval of the Photon Factory Program Advisory
Committee (Proposal No. 2003G149) and under Project No. 2002S2-002 at the
Institute of Material Structure Science, KEK. 
\bibliography{LSFOtex}
\end{document}